\documentclass[a4paper,fleqn,usenatbib]{mnras}

\usepackage[T1]{fontenc}
\usepackage{ae,aecompl}

\usepackage{graphicx}	
\graphicspath{{./Figures/}}
\usepackage{amsmath}	
\usepackage{amssymb}	
\usepackage{subfigmat}
\usepackage{enumerate}
\usepackage{multirow}

\title[Signatures of Protostellar Outbursts]{Observational signatures of outbursting protostars - I: From hydrodynamic simulations to observations}

\author[MacFarlane  et al.]{Benjamin MacFarlane$^{1}$,
Dimitris Stamatellos$^{1}$\thanks{E-mail: dstamatellos@uclan.ac.uk},
Doug Johnstone$^{2,3}$,
Gregory Herczeg$^{4}$,
\newauthor{Giseon Baek$^{5}$,
Huei-Ru Vivien Chen$^{6}$,
Sung-Ju Kang$^{7}$,
Jeong-Eun Lee$^{5}$,
}
\\
\
\\
$^{1}$Jeremiah Horrocks Institute for Mathematics, Physics and Astronomy, University of Central Lancashire, Preston, PR1 2HE, UK\\
$^{2}$NRC Herzberg Astronomy and Astrophysics, 5071 West Saanich Rd, Victoria, BC, V9E 2E7, Canada\\
$^{3}$Department of Physics and Astronomy, University of Victoria, Victoria, BC, V8P 1A1, Canada\\
$^{4}$Kavli Institute for Astronomy and Astrophysics, Peking University, Yiheyuan 5, Haidian Qu, 100871 Beijing, China\\
$^{5}$School of Space Research and Institute of Natural Sciences, Kyung Hee University, 1732 Deogyeong-daero, Giheung-gu, \\
Yongin-si, Gyeonggi-do 446-701, Republic of Korea\\
$^{6}$Department of Physics and Institute of Astronomy, National Tsing Hua University, Taiwan\\
$^{7}$Korea Astronomy and Space Science Institute, 776 Daedeokdae-ro, Yuseong-gu, Daejeon 34055, Republic of Korea\\
}

\date{Accepted XXX. Received 2019; in original form ZZZ}

\pubyear{2019}

\begin{document}
\label{firstpage}
\pagerange{\pageref{firstpage}--\pageref{lastpage}}
\maketitle

\begin{abstract}

Accretion onto protostars may occur in sharp bursts. Accretion bursts during the embedded phase of young protostars are probably most intense, but can only be inferred indirectly through long-wavelength observations. We perform radiative transfer calculations for young stellar objects (YSOs) formed in hydrodynamic simulations to predict the long wavelength, sub-mm and mm, flux responses to episodic accretion events, taking into account heating from the young protostar and from the interstellar radiation field.   We find that the flux increase due to episodic accretion events is more prominent at sub-mm wavelengths than at mm wavelengths; e.g. a factor of $\sim570$ increase in the luminosity of the young protostar leads to a flux increase of  a factor of 47 at 250~\micron\ but only a factor of 10 at 1.3 mm. Heating from the interstellar radiation field may reduce further the flux increase observed at longer wavelengths. We  find that during FU Ori-type outbursts the bolometric temperature and luminosity may incorrectly classify a source as a more evolved YSO, due to a larger fraction of the radiation of the object being emitted at shorter wavelengths. 
\end{abstract}


%
\section{Introduction}

The mass growth of a protostar is expected to occur primarily during the initial phases of its formation while it is still embedded in its parent envelope and  therefore classified as a Class 0/I Young Stellar Object (YSO). The embedded phase, lasting more than $10^{5} \ \text{yr}$ \citep{shu87,evans09,dunham15}, poses observational limitations as radiation from the  protostar is heavily attenuated by the optically thick envelope and reprocessed to far-infrared (IR) through (sub-) mm wavelengths. A common approach to determine the nature of embedded YSOs is to consider their Spectral Energy Distributions (SEDs) and their derived bolometric properties \citep{myersladd93,Andre:2000a,dunham12,Sadavoy:2014a}.

As a protostar accretes gas -either mediated through its protostellar disc or directly from the surrounding envelope- gravitational potential energy is converted to heat on the surface of the protostar and is radiated away. Early theoretical models \citep{shu77,shu87} assumed a constant growth of mass for the protostar. For a solar-mass star, an accretion rate of  $\dot{M} \sim 10^{-5} \ \text{M}_\odot \ \text{yr}^{-1}$ results in a mean luminosity of $\sim 50 \ \text{L}_\odot$ over the course of its protostellar lifetime \citep{stahler80a,stahler80b}. However, observations show that typical YSO luminosities are not that high. The earliest work indicating such a disparity was presented by \citet{kenyon90}, who found that  the mean bolometric luminosity  of embedded protostars is $\sim 1 \ \text{L}_\odot$. More recent surveys with higher numbers of observed YSOs confirm this issue, reffered to as {\it the luminosity problem} \citep{evans09,enoch09,dunham15}. One possible solution to this conundrum is that the protostellar mass growth happens episodically, in short but intense events \citep{Hartmann:1985a, kenyon90,Hartmann:1996a, dunham12,Hartmann:2016a}. Observations of young stars, like the  FUOr-type (after the archetype FU Orionis; \citealp{herbig66,herbig77}) and the EXOr-type (after the archetype EX Lupus; \citealp{herbig89,herbig08}) objects, provide evidence for such behaviour. 

FUOrs are characterised by brightening of up to $5 \ \text{mag}$ \citep{herbig66, herbig77}, and have outburst bolometric luminosities of $\sim \text{few } 100 \ \text{L}_\odot$ \citep{Audard:2014a}. The duration of FUOr-type outbursts may last up to decades - e.g. the archetype FU Orionis has only undergone a dimming of $1 \ \text{mag}$ since its observed $5 \ \text{mag}$ brightening in 1936 \citep{herbig66,herbig77,kenyon00}. The periodicity between episodic accretion events has not yet been constrained, due to the long dimming duration of currently observed FUOrs, and the limited time of the observations. Suggested driving mechanisms for FUOr-type outbursts include external triggers such as binary-disc interactions \citep{reipurth04} and stellar fly-bys \citep{pfalzner08, pfalzner+08}. Additionally, instabilities, taking place within the protostellar disc, have also been widely proposed, e.g. thermal instabilities \citep{bellin95, clarke96, lodato04}, gravitational instabilities \citep{vorobyovbasu05,vorobyovbasu06,vorobyovbasu10a}, and a combination of gravitational instabilities and the  magneto-rotational instability \citep{armitage01, zhu09a, zhu09b, zhu10a, zhu10b, Stamatellos:2011a, Mercer:2017a}.

EXOrs are  less intense,  episodically accreting protostars. These objects are characterised by bolometric luminosities of $1-2 \ \text{L}_\odot$, with outburst luminosities rising to tens of $\text{L}_\odot$ \citep{lorenzetti06,audard10,sipos09,aspin10}. EXOrs' periodicity is on the order of weeks to months \citep{coffey04,audard10}, with duration between accretion events being on the order of $1$ to $2 \ \text{yr}$ \citep{herbig08}.

If FUOr-type outbursts are due to disc gravitational instabilities, one should expect a tendency for FUOr eruptive events to be common in the embedded phase. There is growing evidence that some of the FUOrs are embedded, as shown  by continuum excess at $> 100 \ \mu\text{m}$ \citep{millangabet06}. Recent observations have classified some FUOrs as Class 0/I objects \citep{kospal11,caratti11,Safron:2015a}. Additional support for gravitationally unstable discs triggering FUOrs is provided by the observations of \citet{Liu:2016b} who showed that there are disc asymmetries, possibly spiral arms, in the NIR, suggesting strong disc self-gravity. An embedded FU~Ori-type object is difficult to observe directly as radiation is absorbed very close to the protostar and re-emitted at longer  wavelengths (e.g. OO Serpentis, \citet{kospal07}; NGC 6334l:MM1 \citet{hunter17}; HOPS 383, \citet{Safron:2015a};  EC 53, \citet{yoo17}). 

 \citet{Johnstone:2013a} suggested that variations in the far IR through (sub-)mm emission from protostars might be observed as a proxy for episodic accretion, although they found that there may be a delay in the (sub-)mm response of weeks to months after the outburst has started. Monitoring at (sub-)mm wavelengths could help determine the  characteristics of the outburst (e.g. rise time, magnitude). 
Therefore,  long-term monitoring of star formation regions, at (sub-)mm wavelengths is required. To this end, the James Clerk Maxwell Telescope (JCMT) TRANSIENT Survey is currently undertaking monitoring of eight star-forming regions within $500 \ \text{pc}$ \citep{herczeg17}.  The primary aim of this survey is to observe continuum variability, which may relate to episodic accretion events  in YSOs. The SCUBA2 camera on the JCMT can provide a reliable measure of relative flux brightness changes to $2-3 \%$ for the brightest sources \citep{Mairs:2017a}. Similarly, ALMA is able to observe variability of a few percent at 1.3mm \citep{Francis:2019a}. \cite{Johnstone:2018b} summarizes the results of first 18 months of the JCMT Transient Survey, which has uncovered quasi-periodic variability in the YSO EC 53, with an amplitude of 1.4 at 850~\micron\   \citep{yoo17}, as well as lower amplitude changes from several additional objects over 18 months to 5 years \citep{Mairs:2018a}.

 \citet{kuffmeier18} investigated the flux increase at  wavelengths $20-1000 \ \mu\text{m}$, for heavily embedded YSOs undergoing episodic accretion. Using high resolution hydrodynamic simulations of star-forming regions, \citet{kuffmeier18} modelled the impact of  disc instabilities and environment on episodic accretion. They found that late infall of material can result in episodic accretion, either when infall happens onto the disc (promoting disc gravitational instabilities) or directly onto the protostar. For their highest resolution simulation, the most vigorously accreting star undergoes variations in bolometric luminosity by a factor of $\sim 2$. Taking the inclination-mean flux increase between quiescence and outburst phases, they calculate a factor of $5.5$, $2.6$, $2.0$ and $1.2$ increase at wavelengths of $25$, $70$, $100$ and $1000 \ \mu\text{m}$, respectively. These flux increases happen on a timescale of  decades to centuries. The authors note that although the continuum flux increases are smaller than observed in classical FUOrs, the variability found in their models is larger than currently observed for non-eruptive young protostars \citep{rebull15, flaherty16, rigon17}.  {The variability of their models is also larger than that found by the TRANSIENT Survey \citep{Mairs:2017a, Johnstone:2018b, Mairs:2018a}, } which reports secular changes of order 5-10\% per year  from about 10\% of the protostars bright enough to get good measurements.  If these observed sub-mm secular changes continue for timescales of decades then the magnitude of the flux change will reach  a factor of two or more.

Here, we investigate the impact of episodic accretion on the dust continuum emission of a YSO by carrying out radiative transfer modelling of YSOs produced in  hydrodynamic simulations of cloud collapse that employ episodic accretion. We aim to determine both the change in the bolometric properties and in the continuum flux response to an episodic accretion event. More specifically our goal is to connect the flux increase at various wavelengths with the actual increase in the luminosity of the embedded protostar during an outburst event. { In the follow-up paper (MacFarlane et al. 2019; hereafter {\it Paper II}) we expand on the results of this paper by considering a broader range of protostellar luminosities and structural properties, allowing us to mimic FUOr-type events.}

This paper is structured as follows. In Section~\ref{sec:sims} we describe the hydrodynamic simulations of forming YSOs in collapsing molecular clouds. In Section~\ref{sec:rad_trans} we describe the radiative transfer code that we use to produce SEDs. We then evaluate YSO SEDs, adopting protostellar properties from the hydrodynamic simulations (Section~\ref{sec:hydro_rt}) with or without the influence of heating from an interstellar radiation field. Finally, in Section~\ref{sec:conclusions} we summarise our results.

\section{Radiation Hydrodynamic Simulations}\label{sec:sims}
We use the Smoothed Particle Hydrodynamics (SPH) code  SEREN \citep{hubber11a, hubber11b} to follow the evolution of a pre-stellar core with a total mass of $5.4 \ \text{M}_\odot$ \citep{stamatellos12} that collapses and forms a protostar.  { If we assume a typical star formation efficiency of $\sim 20\%$ \citep[e.g.][]{Andre:2014a} then this cloud will end up forming roughly a solar-mass star}. The initial density of the pre-stellar core follows a Plummer-like profile
\begin{align}
\label{eq:densprofile}
\rho (r) = \frac{\rho_{\text\tiny{C}}}{\left[1+(r/R_{\text\tiny{C}})^{2}\right]^{2}},
\end{align}
where $\rho_{\text\tiny{C}} = 3\times 10^{-18} \ \text{g cm}^{-3}$ is the central density, and $R_{\text\tiny{C}} = 5000 \ \text{AU}$ is the radius at which the density flattens to  $\rho_{\text\tiny{C}}$. The radial extent of the core is $R_\text{\tiny{CORE}} = 50\ 000 \ \text{AU}$, { although most of the mass of the core lies within the inner 10\ 000\, AU}. As per observed values of typical pre-stellar cores \citep[e.g.][]{Andre:2014a}, the initial gas temperature is set to $T = 10 \ \text{K}$. We impose a  random, divergence-free, turbulent velocity field \citep{Bate:2009b}, with power spectrum $P_kdk\propto k^{-4}dk$, to match the scaling laws in molecular clouds \citep{Larson:1981a,Burkert:2000a}, and amplitude such that $\alpha_{_{\rm TURB}}\equiv{U_{_{\rm TURB}}}/{|U_{_{\rm GRAV}}|}=0.3\,$\citep{Jijina:1999a}.
The simulation uses $10^{6}$ SPH particles, such that the mass of each particle is $m_\text{SPH}= 5.4\times10^{-6} \ \text{M}_\odot$. Setting the number of SPH neighbours to $N_\text{\tiny{NEIGH}} = 50$, the minimum resolvable mass is $M_\text{min} \sim m_\text{SPH} N_\text{\tiny{NEIGH}} \sim 3\times10^{-4} \ \text{M}_\odot$. The code uses  an artificial viscosity  to treat shocks, which is time dependent in order to avoid unwanted viscosity away from shocks \citep{morrismonaghan97}.

The radiative transfer method of \citet{stamatellos07} is employed, whereby the SPH particle density, temperature, and gravitational potential are used to estimate the optical depth through which the gas cools and heats \citep[see][]{stamatellos07}. This is then used to solve the energy equation for every time step in the simulation. A pseudo-background radiation field is imposed onto the computational domain that ensures that the gas cannot radiatively cool below a certain temperature,
\begin{align}
T_{\rm BGR} (r) = \left[(10 \ \text{K})^{4} + \frac{L_*}{16 \pi \sigma_\text{\tiny{SB}} r^{2}}\right]^{1/4},
\end{align}
where $L_{*}$ is the protostellar luminosity,  $r$ is the radial distance from the protostar, and $\sigma_\text{\tiny{SB}}$ is the Stefan-Boltzmann constant. The first term on the right hand side of the above equation accounts for a background radiation field with a temperature of $10 \ \text{K}$, and the second term accounts for heating from the young protostar \citep[see][for details]{Stamatellos:2011a, stamatellos12}. The luminosity of the protostar is due to the intrinsic protostellar emission and (mainly) due to accretion of gas onto its surface. This luminosity is parameterised by the protostellar mass $M_{*}$, radius $R_{*}$ and accretion rate $\dot{M_*}$, such that
\begin{equation}\label{eq:p_lum}
L_{*} = \left(\frac{M_{*}}{M_{\sun}}\right)^3 L_{\sun}+L_{\rm acc}\,,
\end{equation}
where
\begin{equation}
L_{\rm acc}=f \ 10.5
\left(\frac{M_*}{M_{\sun}}\right) \left(\frac{\dot{M_*}}{10^{-6} \text{M}_{\odot} \text{yr}^{-1}}\right)
\left(\frac{R_*}{3R_{\sun}}\right)^{-1}L_{\sun}\,,
\end{equation}
and  $f=0.75$ is an efficiency parameter which determines the fraction of the gravitational potential energy radiated away by the protostar, rather than convected into the protostar or being used to drive jest and/or winds \citep{Pudritz:2007a,Offner:2009a,Hartmann:2011b}. 
  The radius of the protostar is set to  $R_{*}=3R_{\odot}$. The effective temperature of the protostar at each computational timestep is computed as 
\begin{equation}
T_{*} = \bigg ( \frac{L_{*}}{4 \pi R_{*}^{2} \sigma_\text{\tiny{SB}}} \bigg )^{1/4}.
\label{eq:p_temp}
\end{equation}

\begin{figure*}
   \centering
   \subfigure{\includegraphics[trim={0.3cm 0.5cm 13cm 0.3cm},clip,width=0.98\columnwidth,keepaspectratio]{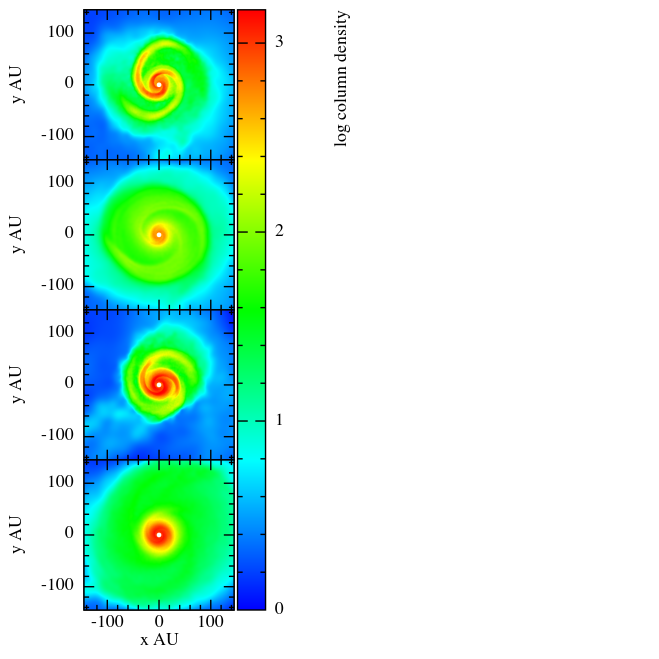}}
   \subfigure{\includegraphics[trim={0.3cm 0.5cm 13cm 0.3cm},clip,width=0.98\columnwidth,keepaspectratio]{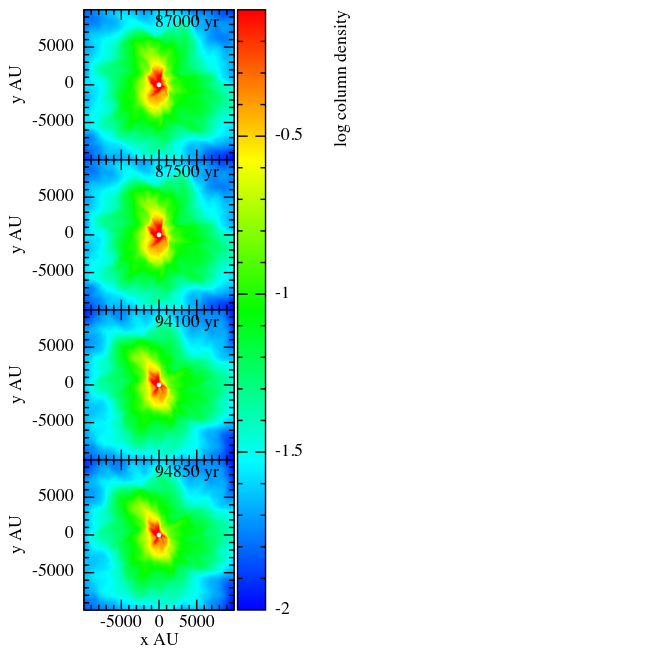}}
\caption{Column density maps (logarithmically scaled; $\text{g cm}^{-2}$) of 4 snapshots during the collapse of a pre-stellar core forming a young protostar and disc. The left column shows the small-scale structure (disc)  and the right column the large-scale structure (envelope) surrounding the YSO. The snapshots correspond to the quiescent and outbursting phases of two events, termed as E1 and E2 (see also Fig.~\ref{fig:temporal}). From top to bottom, the snapshots are the following: E1-quiescent phase, E1-outbursting phase, E2-quiescent phase, E2-outbursting phase. The effect of the outburst on the disc is dramatic (see left column). On the other hand, the large-scale structure of the YSO does not significantly change during the outburst (see right column). }
\label{fig:splash_images}
\end{figure*}

\begin{figure}
\includegraphics[trim={0.3cm 0.3cm 0.3cm 0.3cm},clip,width=0.98\columnwidth, keepaspectratio]{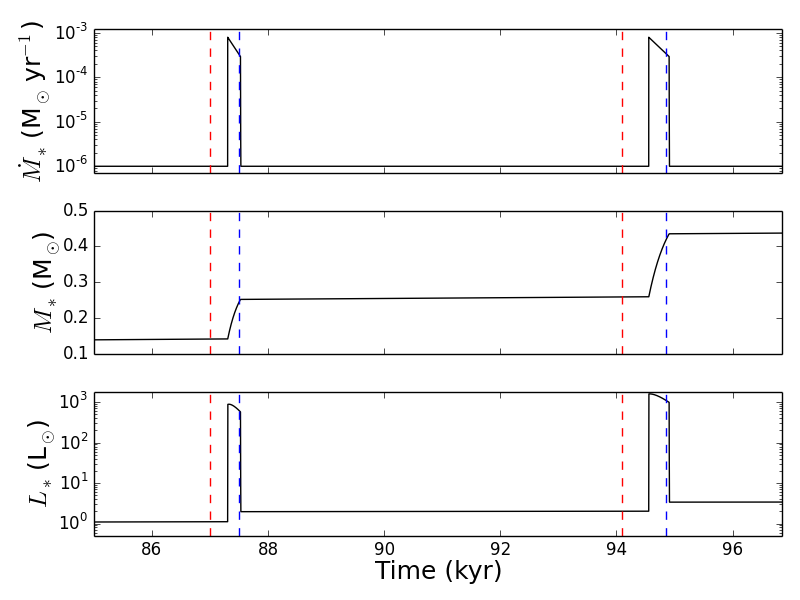}
\caption{Evolution of the protostellar accretion rate $\dot{M}_{*}$ (top), mass $M_{*}$ (middle), and luminosity $L_{*}$ (bottom) for the simulation analysed in this paper. Red and blue dashed lines define the times at which quiescent and outbursting snapshots are selected for detailed radiative transfer calculations. Each pair of red and blue dashed lines correspond to the accretion events E1 and E2, respectively.}
\label{fig:temporal}
\end{figure}

For the simulation analysed in this work, the collapse of the molecular cloud leads to the formation of a  protostar at $\sim 79 \ \text{kyr}$ since the start of the collapse. We assume that a protostar has formed when the density at the centre of the cloud has increased to $10^{-9} {\rm g\ cm^{-3}}$.  The protostar is surrounded by a protostellar disc and an evolving infalling envelope. Accretion of gas onto the protostar occurs episodically \citep[see][]{Stamatellos:2011a,stamatellos12}, resulting in episodic radiative feedback.    Fig.~\ref{fig:splash_images} presents 4 characteristic snapshots from the simulation (2 before and 2 during an outburst). The times of these snapshots are marked in Fig.~\ref{fig:temporal}, which presents the evolution of the protostellar accretion rate, the protostellar mass, and the protostellar luminosity. 

The episodic accretion in this simulation is due to a combination of gravitational instabilities (GI) operating in the outer disc, driving material towards the protostar, and the magneto-rotational instability (MRI) operating in the inner disc region around the young protostar, when the conditions are appropriate for a sufficient ionization fraction  \citep{armitage01, zhu09a, zhu09b, zhu10a, zhu10b, Stamatellos:2011a, Mercer:2017a}. 

 As the disc increases in mass by accreting infalling gas from the envelope it becomes gravitationally unstable, as suggested by the formation of the spiral structures seen in the quiescent phase snapshots in the left column of Figure~\ref{fig:splash_images}. Thus, angular momentum is redistributed outwards efficiently  resulting in  inward  gas flow. The inner disc is hotter and gravitationally stable such that there is no mechanism available to redistribute angular momentum, and material accumulates around the protostar ($\sim 1 \ \text{AU}$). As more gas flows into this region, compressive, viscous, and protostellar heating raise the local temperature further. Once a threshold temperature of $1400 \ \text{K}$ is reached, this region is sufficiently ionised for the MRI to be activated \citep{zhu09a,zhu09b,zhu10a,zhu10b}. The MRI increases the viscosity \citep[the effective viscosity parameter becomes $\alpha_\text{MRI}=0.1$;][]{ss73}, providing a way to redistribute angular momentum and  resulting in rapid gas accretion onto the protostar. During the outburst phase, the accretion luminosity increases dramatically ($\sim10^{3} \ \text{L}_\odot$), heating the disc and disrupting the  spiral structure \citep[outbursting phase snapshots in left column of Fig.~\ref{fig:splash_images}; see also][]{MacFarlane:2017a}. 
 
 Once the gas from the inner region has been accreted, the protostar returns to the quiescent phase. The process of quiescent to outbursting phase repeats multiple times during the evolution of the YSO \citep{stamatellos12}. Outbursts happen every $\sim 1,000$~yr initially, whereas at later stages (but still during the Class O/I stage) every $\sim10,000$~yr \citep[see][]{Stamatellos:2011a,stamatellos12}. This is similar to what is inferred from observations \citep[e.g.][]{Scholz:2013a, HIllenbrand:2015a, Fischer:2019a} \citep[see also][for longer outburst timescale estimates for older, Class I/II, YSOs]{Pena:2019a}. It is assumed that in the quiescent phase gas accretes onto the protostar at a rate of $\dot{M} = 10^{-6} \ \text{M}_{\odot} \text{yr}^{-1}$ (this is a free parameter in our calculation) and the protostellar luminosity is then on the order of $\sim1 \ \text{L}_\odot$. As a comparison, at the same time the accretion rate onto the outer disc is on the order of $\sim 10^{-5}-10^{-4}\ \text{M}_{\odot} \text{yr}^{-1}$. During the outburst phase the accretion onto the protostar is significantly enhanced.
  
  This model is different than the models of  \cite{vorobyovbasu05,vorobyovbasu06,vorobyovbasu10a}. In their models the GIs alone are responsible for the observed outbursts: GIs create clumps in the disc of a young stellar object that eventually fall onto the protostar, resulting in outbursts \citep{dunham13}. { Nevertheless, both models have the same outcome: increased accretion rate onto the young protostar. We expect that the  SED versus accretion luminosity relationship found in this paper should hold in both cases.}

\section{Radiative Transfer Modelling}\label{sec:rad_trans}

\subsection{Radiative Transfer Methods}

The radiative transfer employed within the hydrodynamic simulation uses the 
diffusion approximation method \citep{stamatellos07}. Thus, the computed temperatures are not accurate enough to be compared against observations.  Hence, we post-process the simulations using polychromatic radiative transfer (RT) to model different snapshots in time. 

We employ the 3D radiative transfer code  RADMC-3D\footnote{\href{url}{http://www.ita.uni-heidelberg.de/$\sim$dullemond/software/radmc-3d/} (Last accessed: 07/07/2018)} \citep{dullemond12} that uses the Monte Carlo Radiative Transfer (MCRT) method of \citet{bjorkwood01} to compute the equilibrium temperature for a density distribution, given a set of luminosity sources and a wavelength-dependent opacity \citep[see also][]{Stamatellos:2003c,Stamatellos:2003b,Stamatellos:2004b}. The code works by randomly emitting and subsequently propagating photon packets (henceforth simply referred to as {\it photons})  from each luminosity source within the computational domain. A photon  propagates a distance within the computational domain depending on its randomly sampled optical depth. Once a photon is absorbed, its energy is deposited at that location, and the photon is re-emitted in a random direction.  The method uses a modified random walk algorithm
 \citep{min09, robitaille10} to reduce the computation time required when a photon travels through optically thick regions. Once all photons have propagated through the system, the equilibrium temperature of the system is determined. RADMC-3D then is used to solve the equation of radiative transfer and calculate the wavelength-dependent source function toward an observer location. This method, denoted as Ray-tracing Radiative Transfer (RRT), is used to produce synthetic observations (SEDs and images) of different snapshots of the simulation.

\subsection{Construction of the Radiative Transfer Grid}

RADMC-3D adopts a grid for numerical computations. As such, the SPH density distribution, represented by a large number of SPH particles, must be translated to a grid \citep{Stamatellos:2005b}. To this end, we adopt an adaptive-mesh refinement method.  Initially, we construct a grid cell encapsulating the entire computational domain. This cell is then divided into equal-volume octants.  Each of these sub-cells are then divided into octants and the process continues until each cell contains a user-specified number of SPH particles or reaches the maximum refinement level  (defined by the user). The algorithm adopted here follows that in the RT code {\sc hyperion} \citep{Robitaille:2011a}. 
For the following analyses, the recursive conditions to build the grid are that each grid volume hosts $\leq10$ SPH particles, or a maximum refinement level of $20$ has been reached. These two criteria ensure adequate resolution in the most optically thick regions - namely the protostellar disc and the inner envelope. The computed gas density distribution is then converted to a dust density, adopting a nominal dust-to-gas ratio of 1:100.

Near the  protostar the temperature is high enough for dust to sublimate \citep{lenzuni95,duschl96}. We calculate the dust destruction radius, $R_\text{DUST}$, around the protostar  as 
\begin{align}
R_\text{DUST} = \frac{R_{*}}{2} \bigg( \frac{T_{*} } {T_\text{destr} } \bigg) ^{(4 + \beta_\text{d}) / 2},
\end{align}
by assuming that the dust  heated by the protostar is in thermal equilibrium. $T_{*}$ is computed using Eq. ~\ref{eq:p_temp}, $\beta_\text{d}=3 $ is the dust spectral opacity index ($1 \lesssim \beta_\text{d} \lesssim 3$, see e.g. \citealp{nattatesti04}), and $T_\text{destr} = 1200 \ \text{K}$ is the assumed dust destruction temperature. All cells   interior to  $R_\text{DUST}$ are then set to have zero density.  

\subsection{Dust Opacities}\label{sec:opacities}

\begin{figure}
   \centering
   {\includegraphics[trim={0.3cm 0.3cm 0.5cm 0.3cm},clip,width=0.9\columnwidth]{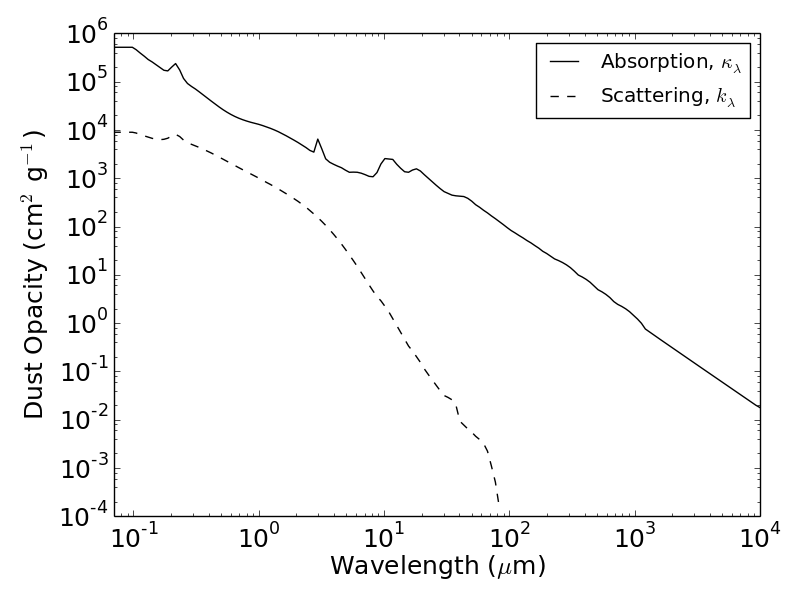}}
\caption{\citet{oh94} (OH5) dust absorption (solid line) and scattering (dashed line) opacities.}
\label{fig:opacities}
\end{figure}

We use the opacities of \citet{oh94} (OH5), for dust grains of a standard MRN mixture (\citealp{mrn77}; $53\%$ silicate, $47\%$ graphite) with thin ice mantles at a density of $10^{6}  \ \text{cm}^{-3}$ (see Fig.~\ref{fig:opacities}). The wavelength range for this opacity table is extended to values of $\lambda < 1 \ \mu\text{m}$ by adopting the optical constants of \citet{drainelee84} for the MRN mixture, computing the absorption and scattering opacities, and rescaling them to match the OH5 values  at $1 \ \mu\text{m}$. 

\subsection{Radiation Sources}

We consider two sources of radiation: (i) the young protostar, and (ii) the interstellar radiation field (ISRF). 

In the RT simulations presented in this paper we  adopt the protostellar luminosity provided by the hydrodynamic simulation (in Paper II,  we treat the protostellar luminosity as a free parameter, so that we can explore a larger parameter space). The emission from the protostar is assumed to follow a blackbody profile at the protostellar temperature. The protostellar radiation is absorbed and re-emitted very close to the protostar; therefore, the blackbody assumption does not affect the results presented in this paper.  

\begin{figure}
   \centering
   {\includegraphics[trim={0.3cm 0.3cm 0.5cm 0.3cm},clip,width=0.9\columnwidth]{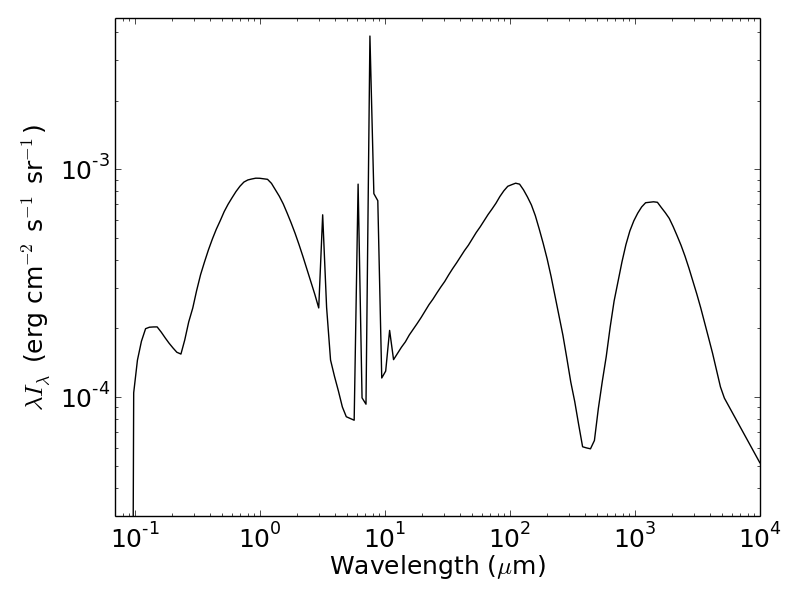}}
\caption{Intensity distribution of the \citet{andre03} interstellar radiation field adopted during the radiative transfer modelling. }
\label{fig:isrf}
\end{figure}

The ISRF (taken into account for a selected set of simulations) considers contributions to the radiation field from the cosmic microwave background, the galactic dust emission, and irradiation from background stars. We adopt the ISRF of \citet{andre03}, which is a modified \citet{black94} ISRF, accounting for emission from poly-aromatic hydrocarbons (Figure~\ref{fig:isrf}). It is assumed that the ISRF isotropically heats from the outside the whole cloud/computational domain (i.e. at $\sim 50,000$~AU), therefore the radiation is somewhat attenuated by the envelope material before it reaches the YSO that we model (i.e. the inner 10,000 AU of the cloud).

\section{Radiative Transfer Modelling of YSOs}\label{sec:hydro_rt}

We focus on two episodic accretion/outburst events in the following modelling and analyses - labelled E1 and E2 respectively (see Fig.~\ref{fig:temporal}). E1 and E2 are both characterised by protostellar luminosity increases of a factor of $\sim 550$, with outbursting phase durations of $\sim200 \ \text{and} \ \sim400\ \text{yr}$, respectively. A simulation snapshot before the outburst phase, and a snapshot during the outburst phase are selected for each of  events E1 and E2. These are denoted by the red (quiescent - at $87.00$ and $94.10 \ \text{kyr}$) and blue (outbursting - at $87.50$ and $94.85 \ \text{kyr}$) dashed lines of Fig.~\ref{fig:temporal}, respectively. These snapshots correspond to the column density maps presented in Fig.~\ref{fig:splash_images}.  { The outburst snapshots for both events were chosen near the end of each outburst so that the region around the protostar has time to adjust itself to the additional heating provided during the outburst \citep{MacFarlane:2017a}.}

MCRT is carried out on each of these snapshots. We examine two models, (i) including and (ii) excluding the contribution from the ISRF. For the MCRT computations  $10^{6}$ photon packets are used to represent the radiation from the protostar and $10^{7}$ photon packets for the ISRF. These values ensure that the statistical noise when calculating the equilibrium temperature is similar for both types of luminosity sources.

\subsection{Temperature Distributions}
\label{sec:temp}

\begin{figure*}
   \centering
   \subfigure{\includegraphics[trim={0.2cm 0.5cm 1.2cm 0.4cm},clip,width=0.98\columnwidth]{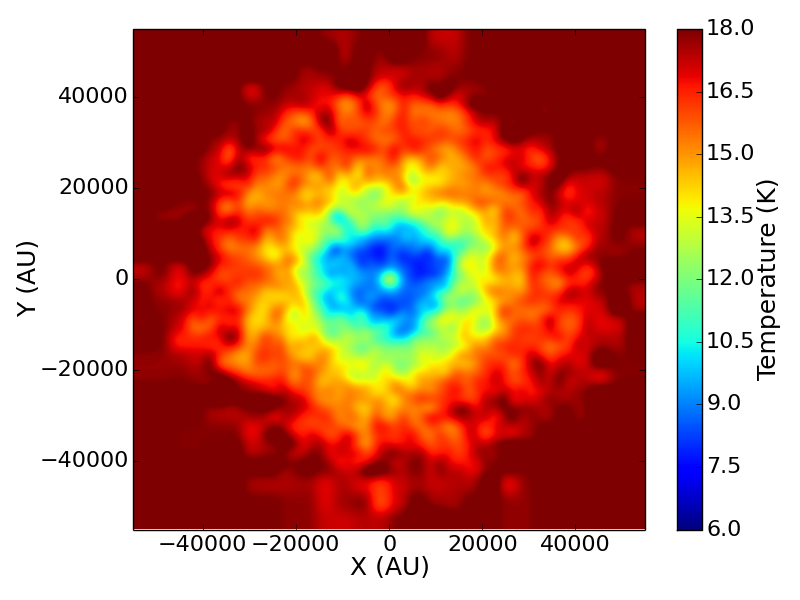}}
   \subfigure{\includegraphics[trim={0.2cm 0.5cm 1.2cm 0.4cm},clip,width=0.98\columnwidth]{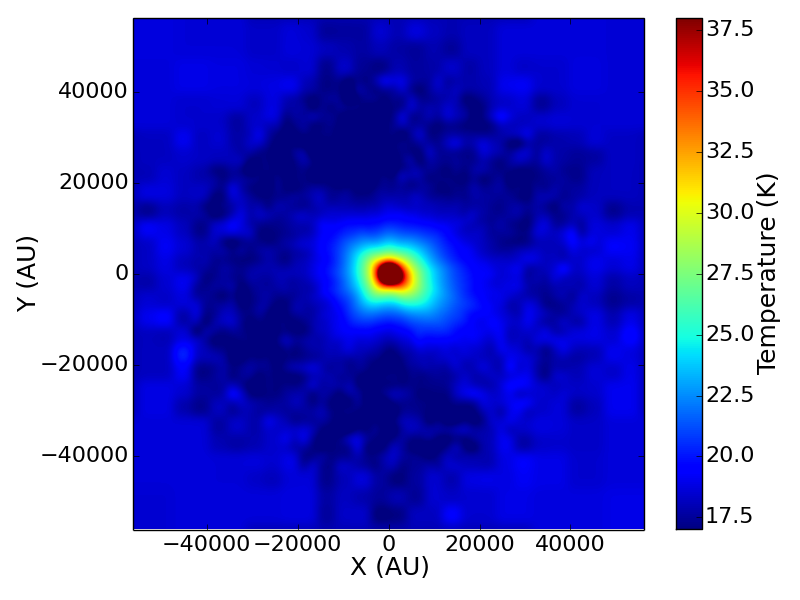}}
   \subfigure{\includegraphics[trim={0.2cm 0.5cm 1.2cm 0.4cm},clip,width=0.98\columnwidth]{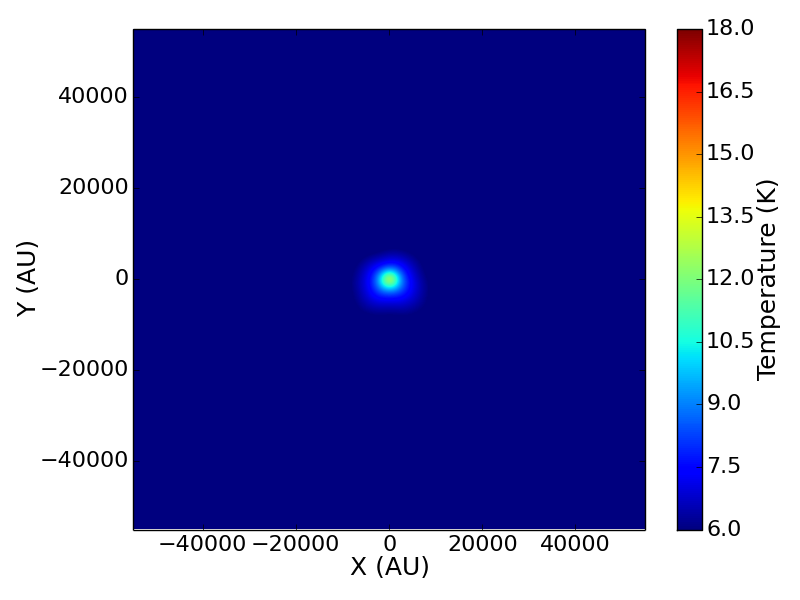}}
   \subfigure{\includegraphics[trim={0.2cm 0.5cm 1.2cm 0.4cm},clip,width=0.98\columnwidth]{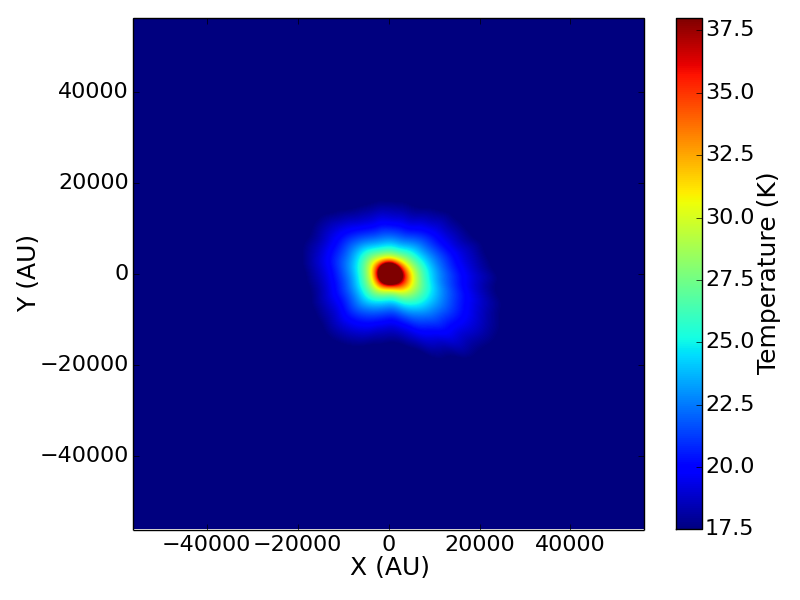}}
\caption{Temperature distributions on the $x-y$ plane ($z=0 \ \text{AU}$; in Kelvin), for radiative transfer models of event E1. Left and right panels represent quiescent and outbursting phases, respectively,  while top and bottom panels represent MCRT calculations with and without ISRF contributions, respectively.}
\label{fig:rt_properties}
\end{figure*}

The $x-y$ temperature maps (taken at $z=0 \ \text{AU}$, i.e. at the disc midplane) for event E1 are presented in Fig.~\ref{fig:rt_properties}. Left and right panels represent the temperature profiles for quiescent and outbursting snapshots, respectively. The top/bottom panels represent MCRT computations with/without heating contributions from the ISRF. 

For MCRT models that include  ISRF heating, the temperature reaches a minimum within the envelope, at a radius determined by the relative strength of the ISRF  with respect to the protostellar luminosity. The radius of the temperature minima for the quiescent phase snapshot we present in Fig.~\ref{fig:rt_properties} (top left)  is $\sim12,000 \ \text{AU}$ ($\sim 5$~K), whereas for the outburst phase (top right) it is
 $\sim23,000 \ \text{AU}$ ($\sim 15$~K).
 
For MCRT models where the ISRF heating is neglected, the temperature decreases monotonically with distance from the central protostar. The temperature minima, therefore, occur at the  outer edge of the envelope ($\sim 50,000 \ \text{AU}$): the minimum temperature is  $\sim 5$~K  for the quiescent (Fig.~\ref{fig:rt_properties}, bottom left) and $\sim 15$~K for the outbursting model (Fig.~\ref{fig:rt_properties}, bottom right).

\subsection{Spectral Energy Distributions}\label{sec:sed}
We use the calculated temperature distributions to produce SEDs for each snapshot. For these calculations in RADMC-3D, we take into account only the radiation that comes from the region $R \leq 10 \ 000 \ \text{AU}$ around the protostar. We therefore assume that any background emission (i.e. any emission outside $10 \ 000 \ \text{AU}$) is filtered out either directly  or removed by considering external annuli and subtracting to only get the flux associated with the source. { Our observed region is therefore roughly the typical size of a Class 0 object as mapped in the sub-mm (e.g. at 850~\micron\ by JCMT). Considering the distances of nearby star-forming regions where such objects are observed ($150-500$~pc) and the resolution of the JCMT at 850~\micron\ (14.\arcsec6) then the spatial resolution  is $\sim2200-7300$~AU, which provides a lower limit on the integration area. In Paper II we also discuss integration areas of 1000~AU that can be probed by higher resolution observations, e.g. by ALMA.} We adopt a distance to the source of $d = 140 \ \text{pc}$ and an inclination of $i = 0^\circ$. We define $i = 0^\circ$ to be the orientation such that the protostellar disc of the YSO is face-on.

\begin{figure*}
   \centering
   \subfigure{\includegraphics[trim={0.8cm 0.8cm 0.1cm 0.5cm},width=0.95\columnwidth,keepaspectratio]{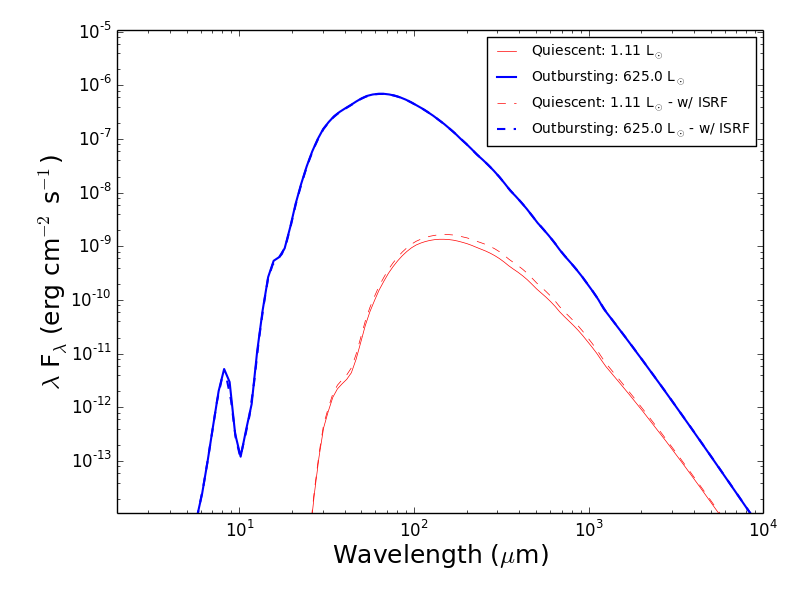}}
   \subfigure{\includegraphics[trim={0.8cm 0.8cm 0.3cm 0.5cm},width=0.95\columnwidth,keepaspectratio]{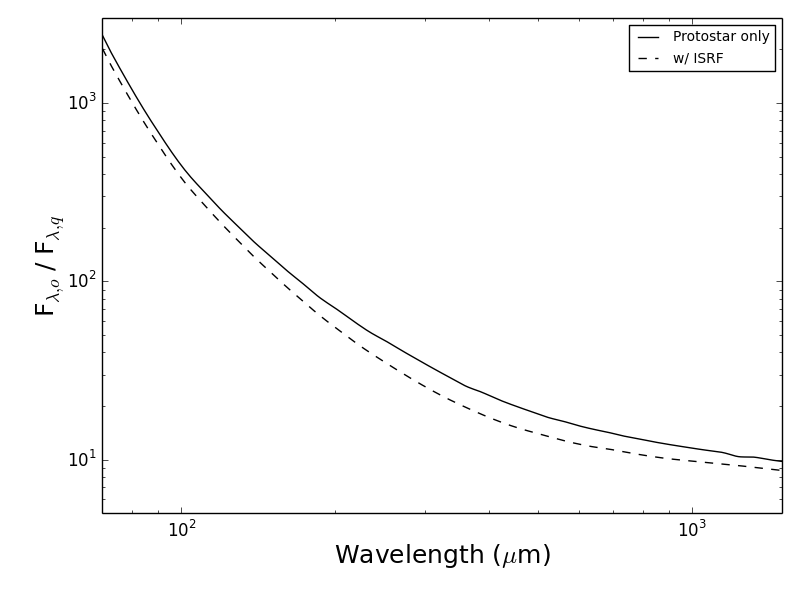}}
   \subfigure{\includegraphics[trim={0.8cm 0.8cm 0.1cm 0.5cm},width=0.95\columnwidth,keepaspectratio]{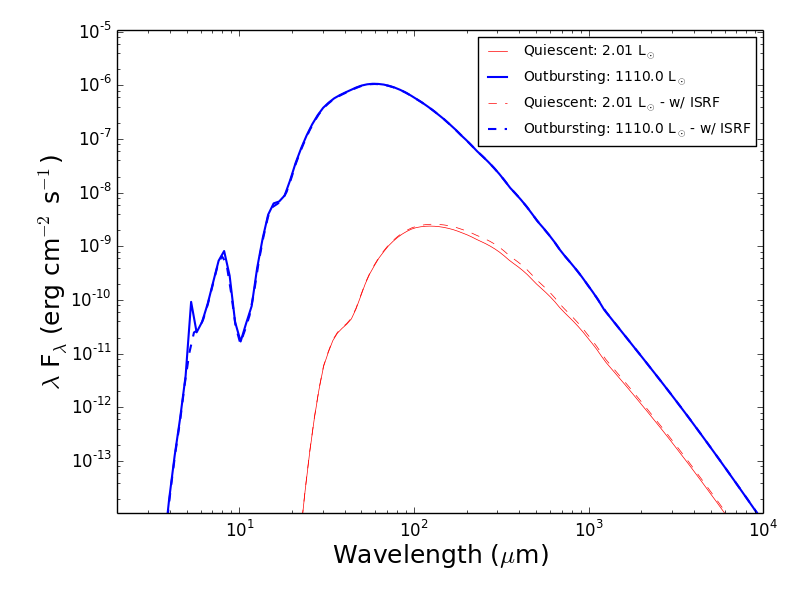}}
   \subfigure{\includegraphics[trim={0.8cm 0.8cm 0.3cm 0.5cm},width=0.95\columnwidth,keepaspectratio]{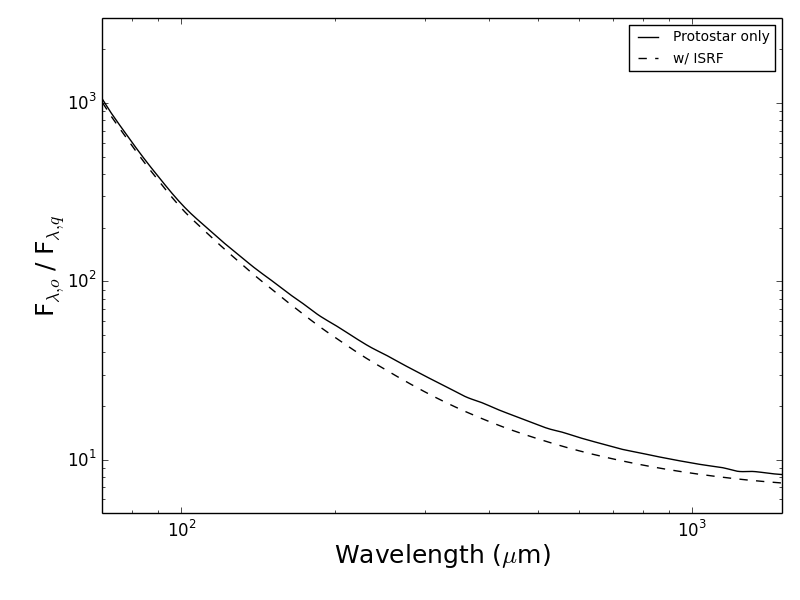}}
\caption{Spectral Energy Distributions (left) and the ratios of outbursting-to-quiescent flux (right), with and without heating from the ISRF, for event E1 (top) and E2 (bottom). On the left graphs, red and blue lines represent SEDs for quiescent and outbursting phase snapshots, respectively. Solid/dashed lines  correspond to RT models without/with the ISRF (as indicated on each graph).}
\label{fig:seds}
\end{figure*}

We present SEDs for all snapshots and luminosity source combinations in Fig.~\ref{fig:seds}. Top and bottom panels represent SEDs for events E1 and E2, with red/blue lines indicating the emission from quiescent/outbursting phase. Dashed/solid lines represent temperature distributions computed with/without ISRF contributions. On the right column of  Fig.~\ref{fig:seds},  we present the ratio between the outburst and the quiescent phase flux over a narrower wavelength range to better illustrate the differences between the models with and without ISRF.

\subsection{Flux Increases in Episodically Accreting Protostars}\label{sec:flux_ratios}

 The ratio of outbursting to quiescent flux at long wavelengths serves as a proxy for the change in protostellar luminosity during an accretion event. In Table~\ref{tab:ratios} we present the outbursting and quiscent luminosities for each event. We also present the flux ratios between outbursting and quiescent snapshots (see SEDs of Fig.~\ref{fig:seds}). We compute ratios at wavelengths corresponding to continuum bands of the {\it Herschel Space Observatory}, JCMT and the Atacama Large (Sub-) Millimetre Array (ALMA) ($70, \ 250, \ 350, \ 450, \ 850 \ \text{and} \ 1300 \ \mu\text{m}$).

\begin{table*}
	\caption{Ratio of fluxes at various wavelengths (outbursting flux. $F_{\lambda\text{,o}}$, divided by quiescent flux, $F_{\lambda\text{,q}}$). Ratios are calculated for two accretion events (E1, E2) using the source luminosities adopted in the MCRT models (with/without ISRF). $M_*$ is the mass of the protostar, $L^*$  is its luminosity (quiescent and outbursting), and $L^*_{\text{o}} / L^*_{\text{q}}$ is the ratio of outbursting to quiescent luminosity for each event.}
	\renewcommand{\arraystretch}{1.3} 
	\begin{tabular}{ c c c c c c c c c c c  }
		\hline
		Accretion &  $M_*$ (M$_{\sun}$) & $L^*$ (L$_{\sun}$) &$L^*_{\text{o}} / L^*_{\text{q}}$ & Luminosity Source(s) & \multicolumn{6}{c}{$F_{\lambda\text{,o}} / F_{\lambda\text{,q}}$}\\
		 Event & & & & & 70 $\mu$m & 250 $\mu$m & 350 $\mu$m & 450 $\mu$m & 850 $\mu$m & 1300 $\mu$m \\
		\hline
		\multirow{2}{*}{E1} & 0.142& 1.1 & \multirow{2} {*}{570} & Protostar &  $2400$ & $47$ & $27$ & $20$ & $12$& $10$\\
		& 0.254& 625 &  &Protostar+ISRF &  2000 & 35 & $21$ & $15$ & $10$& $9.1$\\
		\multirow{2}{*}{E2} &0.258 &2.0 & \multirow{2}{*}{555} & Protostar &  $1000$ & $39$ & $24$ & $18$ & $10$& $8.6$\\
		&0.417 &1110 & &Protostar+ISRF &  1000 & 32 & $19$ & $15$ & $9.0$& $7.7$\\
		\hline
	\end{tabular}
\label{tab:ratios}
\end{table*}

In the quiescent phase SEDs, very little flux is emitted at wavelengths shorter than $\lambda < 100 \ \mu \text{m}$. In contrast, the outburst phase SEDs exhibit significant increases in flux, due to the enhanced heating of the envelope. This result suggests that an embedded source that was previously undetectable, may be observed at  $\lambda < 100 \ \mu \text{m}$ (e.g. with the upcoming James Webb Space Telescope; \citealp{gardner06}) if the protostar undergoes an outburst.

The magnitude of the flux ratios in our models are much higher than those of EC 53, an embedded Class I YSO. EC 53 underwent an increase in $850 \ \mu\text{m}$ flux by a factor $1.5$ \citep{yoo17}, indicative of a factor of $\sim4$ increase in accretion luminosity.  In contrast, the protostar in our hydrodynamic simulation undergoes luminosity increases of order $\sim 550$, leading to flux increases at 850 \micron\ an order of magnitude higher than observed in EC 53.

\begin{figure}
   \centering
   {\includegraphics[trim={0.8cm 0.8cm 0.3cm 0.5cm},clip,width=0.9\columnwidth]{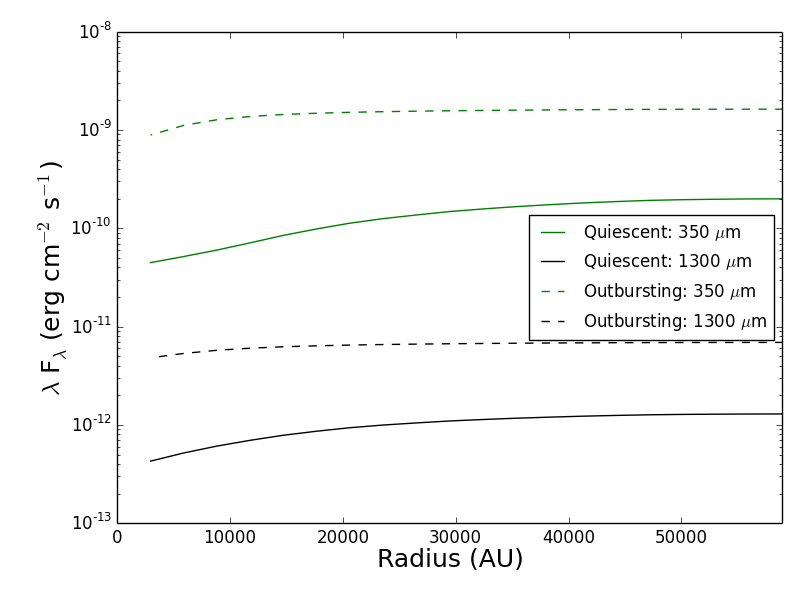}}
\caption{$\lambda F_\lambda$ at $350 \ \mu\text{m}$ (green) and $1300 \ \mu\text{m}$ (black) for E1 event snapshots, as a function of radius within which radiation is included. Solid lines represent the quiescent phase of event E1, with the outbursting phase snapshot represented by dashed lines. RT calculations for both cases are carried out with both protostellar and ISRF heating.}
\label{fig:lamflam_beam}
\end{figure}

\subsection{Impact of the ISRF on the SEDs}\label{sec:isrf_impact}

We see in Fig.~\ref{fig:seds} that heating from the ISRF  results in an increase in flux at mid-IR and longer wavelengths, but only 
during the quiescent phase of the protostar. In contrast, the outbursting phase SEDs show no discernible change in long wavelength flux when the ISRF heating is included. 
As noted in Section~\ref{sec:temp}, the heating of the envelope is dominated by the ISRF outside $\sim12,000 \ \text{AU}$ for the quiescent phase and  outside $\sim23,000 \ \text{AU}$ for the outburst phase of event E1. As in both cases the regions in which the ISRF dominates is outside the region  over which the SED is computed ($10,000 \ \text{AU}$), only small differences in long wavelength emission are expected. { For a larger integration area (say $20,000$~AU) more significant (but still rather small) differences are expected during the quiescent phase.}

Investigating this further, in Fig.~\ref{fig:lamflam_beam} we present $\lambda F_\lambda$  as a function of the  size of the region around the protostar from which we calculate the flux for event E1. Values are computed at wavelengths of $350 \ \text{and} \ 1300 \ \mu\text{m}$, including radiation from within an increasing radius in the models that include heating from the ISRF.  In the quiescent snapshot (solid lines), there is a noticeable drop in long-wavelength emission with decreasing radius. This is  because the flux from the outer envelope, which is heated by the ISRF, is not included. On the other hand, in the outbursting phase heating of the envelope is due primarily to the protostellar luminosity and  the ISRF contribution  is minimal. As a result, there is no significant change in long wavelength emission with decreasing distance from the protostar. This confirms that the relative strengths of the luminosities from protostar and the ISRF determine whether significant long wavelength flux increases are observed, when heating from an ISRF is present.

 In the case of  strong heating from the ISRF and a relatively small protostellar luminosity increase (as, e.g., in EXOr-type objects), the above effect could lead to just a slight increase in the radiation at long wavelengths if the flux is calculated over a large integration area (i.e. due to low angular resolution). To mitigate this issue, ideally one should conduct high-resolution observations of the  central regions of the YSO (e.g. $\stackrel{<}{_\sim}10,000-20,000$~AU) ensuring that radiation is received from regions where heating is predominantly due to protostellar radiation and not from the ISRF.

\subsection{The effect of inclination on the SEDs}\label{sec:inclinations} 

\begin{figure}
   \centering
   {\includegraphics[trim={0.8cm 0.8cm 0.5cm 0.5cm},clip,width=0.9\columnwidth]{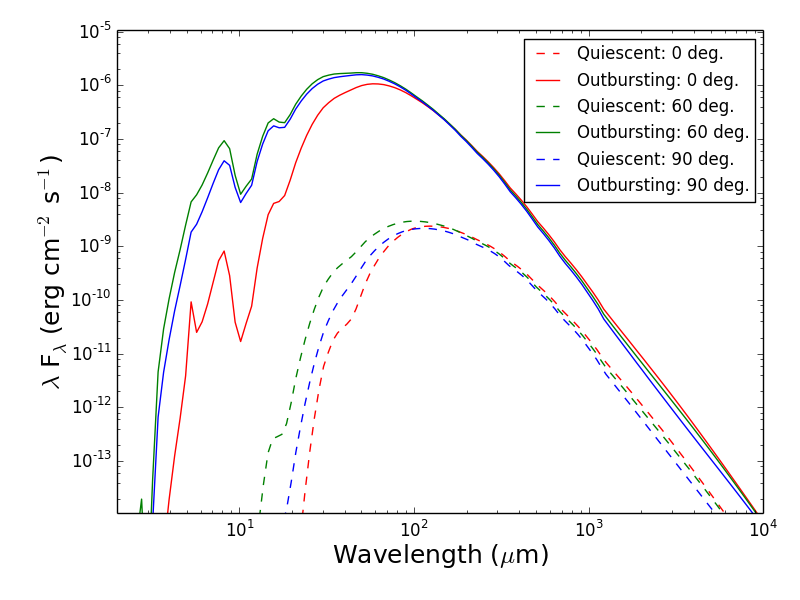}}
\caption{SEDs for event E2 (computed with both the protostellar and ISRF heating), computed at inclinations of $0^\circ$ (red), $60^\circ$ (green) and $90^\circ$ (blue). Quiescent phase SEDs are represented by dashed lines, while solid lines represent the outbursting phase.}
\label{fig:sed_inclin}
\end{figure}

The observed SEDs of YSOs depend on their orientation.  In Fig.~\ref{fig:sed_inclin} we present SEDs for event E2 (with ISRF heating), at inclinations of $0^\circ, \ 60^\circ \ \text{and} \ 90^\circ$. An inclination of $i = 0^\circ$ denotes observing onto the $x-y$ plane (disc face-on), whereas an inclination of $i = 90^\circ$ denotes observing onto the $x-z$ plane (disc edge-on). As mentioned earlier (see Fig.~\ref{fig:splash_images}),  the simulated pre-stellar collapsing cloud evolves to a  highly asymmetric state.  Fig.~\ref{fig:sed_inclin} demonstrates that for both quiescent and outbursting phases, asymmetries in the envelope lead to a non-monotonic relationship between flux and inclination for $\lambda < 200 \ \mu\text{m}$. This is in contrast
to models of Class 0/I YSOs with axisymmetric envelopes flattened due to rotation, in which the increase of inclination from $i = 0^\circ$  to $i = 90^\circ$  leads to decreasing flux at all wavelengths \citep[e.g.][]{whitney03a,whitney03b}. At longer wavelengths (e.g. 1.3~mm), the flux decreases with increasing inclination in both set of models.

Radiation at  $\lambda < 200 \ \mu\text{m}$  is  attenuated due to the relatively high dust opacity, so that the radiation received by the observer is from the outer layers of the envelope.   The SEDs presented  imply that at $60^\circ$   the radiation comes from deeper  within the envelope than at $0^\circ$ or  $90^\circ$. This radiation comes from the inner,  hotter and more dense envelope at $i = 60^\circ$, compared to $i = 0^\circ$ $i = 90^\circ$.  On the other hand, radiation  at $\lambda > 200 \ \mu\text{m}$, comes from  almost the entire YSO (i.e. even from the inner parts of the envelope), and it decreases slightly with increasing inclination.

\begin{figure}
   \centering
   \subfigure{\includegraphics[trim={0.5cm 0.5cm 0.5cm 0.3cm},clip,width=0.9\columnwidth,keepaspectratio]{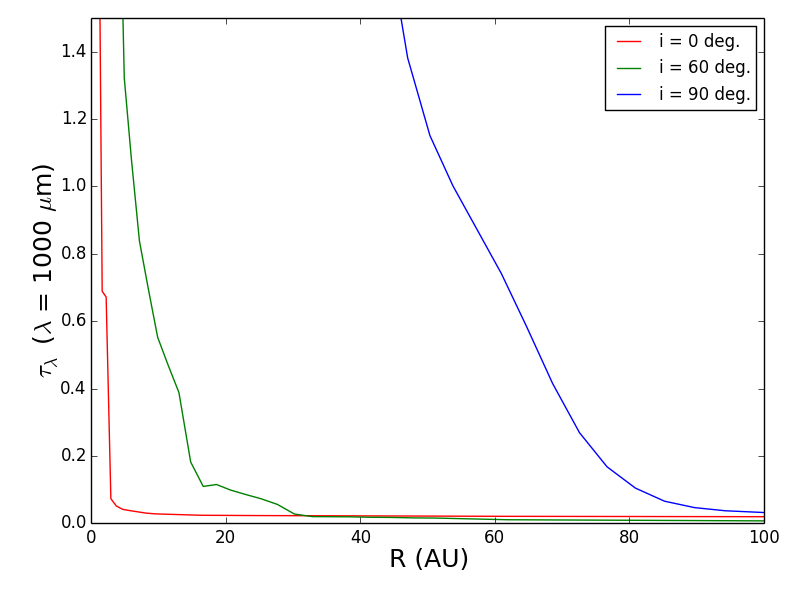}}
   \subfigure{\includegraphics[trim={0.5cm 0.5cm 0.5cm 0.3cm},clip,width=0.9\columnwidth,keepaspectratio]{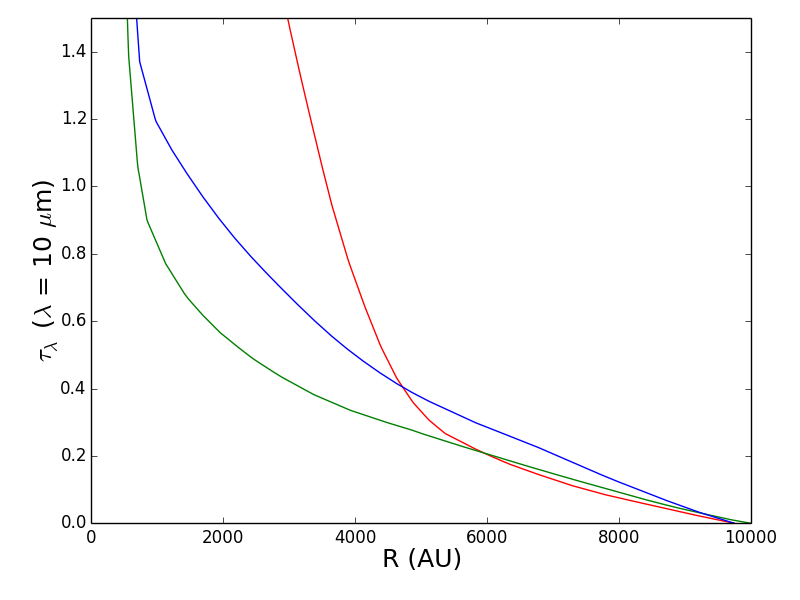}}
\caption{Optical depth from the observer to a specific radius  $R$ away from the protostar, at different viewing angles, for $\lambda =1000 \mu\text{m}$ (top) and $\lambda=10 \mu\text{m}$ (bottom). In each panel, red, green and blue lines denote an  inclination of $0^\circ,  60^\circ$ and $90^\circ$, respectively.}
\label{fig:tau_r}
\end{figure}

Investigating further, in Fig.~\ref{fig:tau_r} we present radial optical depth profiles, for $\lambda = 10 \ \mu\text{m}$ and $\lambda = 1000 \ \mu\text{m}$. This calculation is carried out for the quiescent phase snapshot of E2 (the SEDs of which are presented in Fig.~\ref{fig:sed_inclin}), calculating the optical depth from the observer toward the central protostar. The optical depth, $\tau_\lambda (R)$, is calculated by numerically integrating the density from  $R_\text{OUT}=10 \ 000 \ \text{AU}$, i.e. at the radial range over which the SEDs of Fig.~\ref{fig:sed_inclin} are computed, to a given radius $R$, i.e.
\begin{equation}\label{eq:tau}
\tau_\lambda (R) = - \kappa_\lambda \int^{R}_{R_\text{OUT}} \rho(r) \ \text{d}r, 
\end{equation}
where $\kappa_\lambda$ is the dust opacity. { As the outer density of the cloud drops approximately as $r^{-4}$ (see Equation~\ref{eq:densprofile}), the contribution to the optical depth from dust outside 10,000~AU at these long wavelengths is minimal.}
The top panel of Fig.~\ref{fig:tau_r} shows that at $\lambda = 1000 \ \mu\text{m}$ radiation comes progressively from deeper within the envelope as the inclination increases (as indicated by the radius at which $\tau_\lambda \sim 1$). In the case of $\lambda = 10 \ \mu\text{m}$ (left panel of Fig.~\ref{fig:tau_r}),  the radius at which $\tau_\lambda \sim 1$ is farther away from the protostar and does not decrease monotonically as the inclination increases. In this case,  the radius at which $\tau_\lambda = 1$ is smaller for  $i = 60^\circ$, and larger for $i = 0^\circ$, confirming  the inclination variation of the SEDs seen in Fig.~\ref{fig:sed_inclin}. Therefore, in the non-axisymmetric system that we model,  the use of the  inclination angle to interpret an SED, is not particularly useful.

\subsection{YSO Classification}\label{sec:class}

Each of the model snapshots presented here represent a deeply embedded protostar, i.e. a Class 0 YSO. To classify a YSO, the bolometric temperature, and the sub-mm to bolometric luminosity ratio may be used. Sources with a bolometric temperature of  $T_\text{BOL} < 70 \ \text{K}$ are classified as Class 0 YSOs \citep{chen95}. The bolometric temperature is calculated from the equivalent temperature of a blackbody with the same flux-weighted mean frequency as the YSO. Alternatively, Class 0 YSO sources are identified as those with a ratio of sub-mm to bolometric luminosity, $L_\text{sub-mm}/L_\text{BOL} > 0.5 \%$ \citep{andre93}. The sub-mm luminosity is defined as the integrated flux for wavelengths longer than $350 \ \mu \text{m}$.

The ratio of the sub-mm to bolometric luminosity and the bolometric temperature are numerically calculated  (see Table~\ref{tab:ysoclass}) from the SED, 
\begin{equation}\label{eq:lum_ratio}
\dfrac{L_\text{sub-mm}}{L_\text{BOL}} = \dfrac{ \int^{\nu(350\mu\text{m})}_{0} F_\nu \ \text{d}\nu}{ \int^{\infty}_{0}F_\nu \ \text{d}\nu},
\end{equation}
and
\begin{equation}
T_\text{BOL} = 1.25\times10^{-11} \dfrac{ \int^{\infty}_{0} \nu F_\nu \ \text{d}\nu}{ \int^{\infty}_{0}F_\nu \ \text{d}\nu}\ {\rm {Hz^{-1}\ K}}\,,
\end{equation}
 \citep[see][]{myersladd93}.

\begin{table}
	\caption{The bolometric luminosity ($L_\text{BOL}$), bolometric temperature ($T_\text{BOL}$), and submm to bolometric luminosity ratio ($L_\text{submm}/L_\text{BOL}$) for events E1 and E2 (before and during an outburst), for an inclination of $i=0^\circ$.}
	\renewcommand{\arraystretch}{1.3} 
	\begin{tabular}{ccccc}
		\hline
		Accretion& Luminosity & $L_\text{BOL}$  & $T_\text{BOL}$ & L-ratio\\
		 Event & Source(s) &  ($\text{L}_{\odot}$) & ($\text{K}$) & (\%) \\
		\hline
		\multirow{2}{*}{E1-Quiescent} & Protostar & $1.1$ & $25$ & $8.9$\\
		&Protostar+ISRF & $1.3$ & $25$ & $9.2$\\
		\multirow{2}{*}{E1-Outbursting} & Protostar & $500$ & $59$ & $0.38$\\
		&Protostar+ISRF & $500$ & $58$ & $0.38$\\
		\multirow{2}{*}{E2-Quiescent} & Protostar & $1.9$ & $29$ & $6.2$\\
		&Protostar+ISRF & $2.1$ & $28$ & $6.7$\\
		\multirow{2}{*}{E2-Outbursting} & Protostar & $790$ & $66$ & $0.26$\\
		&Protostar+ISRF & $790$ & $65$ & $0.26$\\
		\hline
	\end{tabular}
\label{tab:ysoclass}
\end{table}

When using the bolometric temperature, all stages of the evolution of the YSO presented in Table~\ref{tab:ysoclass} are  classified as Class 0 (note, however, that there is also a dependence on inclination, see Paper II). In contrast, when using the ratio of sub-mm to bolometric luminosity, the YSO is classified as a Class 0 object only during its quiescent phase. In the outbursting phase the ratio is  $< 0.5\%$, indicative of later-phase YSOs \citep{andre93}. This result is due to a significant fraction of the protostellar radiation in the outbursting phase being emitted at  shorter wavelengths, causing the sub-mm to bolometric luminosity ratio to decrease significantly. As the envelope is more massive than the protostar itself, the YSO is still a Class 0 source \citep{andre03}. This implies that using the ratio of sub-mm to bolometric luminosity may lead to an incorrect classification for  very luminous FUOrs. Heating from the ISRF has little significant effect on the computed bolometric values and therefore on the classification (see Table~\ref{tab:ysoclass}).

{ Comparing the $L_\text{BOL}$ values we calculate to the observed luminosity distribution of YSOs, we find that the quiescent phase values are comparable to the observed mean values for Class 0/I YSOs ($\sim 1 \ \text{L}_\odot$; \citealp{evans09,dunham13}).  The bolometric luminosities  in the outbursting phase are higher than the typical $L_\text{BOL}$ of FU Ori objects \cite[$100-300\text{L}_\odot$;][]{Audard:2014a}. However, there are a few observations of FU Ori objects with such high luminosities, e.g. Z CMa \cite[$400-600\text{L}_\odot$;][]{Hartmann:1996a} and  V1057 Cyg  \cite[$250-800\text{L}_\odot$;][]{Hartmann:1996a}. This indicates that the FU Ori-type objects that we model here represent rather rare  events. Quick evolution within the Class 0 phase (e.g. \citealp{dunham15}), and the relatively short duration of outbursts suggests that observations of such luminous YSOs would be rare.}

\section{Conclusions}\label{sec:conclusions}

In this paper, we presented polychromatic radiative transfer simulations of episodically outbursting YSOs formed in hydrodynamic simulations. The primary aim of this work has been to determine the changes in the continuum flux at various wavelengths and connect these with the luminosity increase of the young protostar. Additionally, we explored the change in the bolometric properties of YSOs due to outburst events. 

The radiative transfer models that we presented utilized protostellar properties directly from hydrodynamic simulations that model YSOs undergoing FUOr-type outbursts { (see Paper II for a parameter study)}. SEDs have been computed both in quiescent and outbursting phases for two episodic accretion events. The integrated flux has been computed over the central $R \leq 10 \ 000 \ \text{AU}$ region of each YSO, at various inclinations. The flux increase due to episodic accretion events is more prominent at NIR wavelengths (a factor of up to 50 increase at 250~\micron) than at mm wavelengths (a factor of 10 increase at 1.3 mm). The effect of heating from the ISRF may somewhat attenuate the expected flux increase during an episodic accretion event. The reduced flux contrast between quiescent and outbursting phase may be avoided if resolution permits the monitoring of the inner regions of the YSO, where the heating happens primarily due to protostellar radiation { (see Paper II)}.  

We have also calculated bolometric luminosities and temperatures before and during outbursts and found that a correct classification as a Class 0 object is generally made when using the bolometric temperature as a proxy for evolutionary phase. However, when using the ratio of sub-mm to bolometric luminosity  during outbursts the YSO is often incorrectly classified as Class I, indicating that this criterion is not an optimal indicator of the evolutionary phase for outbursting YSOs.

{ The work presented in this paper describes the observational characteristics of young Class 0/I protostars undergoing episodic mass accretion and therefore large luminosity outbursts, akin to FU Ori-type objects. In the follow-up paper (Paper II) we consider a  range of protostellar luminosity and  YSO structure and we find that the flux increases at different wavelengths also depend on the specific density structure and the geometry of the young protostar.  We provide diagnostics to infer the luminosity of episodically outbursting embedded protostars using observations at FIR and mm wavelengths.} 

\section*{Acknowledgements}
BM is supported by STFC grant ST/N504014/1. DS is partly supported by STFC grant ST/M000877/1. DJ is supported by the National Research Council Canada and by an NSERC Discovery Grant.
GH is supported by general grant 11773002 awarded by the National Science Foundation of China. Simulations were performed using the UCLAN HPC facility and the COSMOS Shared Memory system at DAMTP, University of Cambridge operated on behalf of the STFC DiRAC HPC Facility. This equipment is funded by BIS National E-infrastructure capital grant ST/J005673/1 and STFC grants ST/H008586/1, ST/K00333X/1. {\sc Seren} has been developed and maintained by David Hubber, who we thank for his help. 
Column density maps were generated using the visualization software SPLASH \citep{price07}. This work is supported by the JCMT-Transient Team.
%


\bibliographystyle{mnras}

\bibliography{macfarlane18i} 



\bsp	
\label{lastpage}
\end{document}